\documentclass[review]{elsarticle}

\usepackage{amssymb}
\usepackage{url}
\usepackage{amsfonts,amsmath}
\usepackage{graphicx}
\usepackage{stfloats}

\usepackage[numbers]{natbib}
\usepackage{algorithmic}
\usepackage[ruled,vlined]{algorithm2e}
\usepackage{amsmath, bm}
\usepackage{subcaption}
\usepackage{caption}

\usepackage{multirow}
\usepackage{xfrac}
\usepackage{color}
\usepackage{array}
\usepackage[table]{xcolor}

\usepackage[Symbol]{upgreek}

\usepackage[T1]{fontenc}

\journal{arXiv.org}

\begin{document}

\begin{frontmatter}

\title{Hardware Architecture Proposal for TEDA algorithm to Data Streaming Anomaly Detection}

\author[1,2]{Lucileide M. D. da Silva}
\ead{lucileide.dantas@ifrn.edu.br}

\author[1]{Maria G. F. Coutinho}
\ead{gracielly@dca.ufrn.br}

\author[1]{Carlos E. B. Santos}
\ead{ceduardobsantos@gmail.com}

\author[4]{Mailson R. Santos}
\ead{mailsonribeiro@ufrn.edu.br}

\author[4]{Luiz Affonso Guedes}
\ead{affonso@dca.ufrn.br}

\author[3]{M. Dolores Ruiz}
\ead{mariloruiz@ugr.es}

\author[1,4]{Marcelo A. C. Fernandes\corref{cor1}\fnref{ca1}}
\ead{mfernandes@dca.ufrn.br}

\cortext[cor1]{Corresponding author}
\fntext[ca1]{Present address: John A. Paulson School of Engineering and Applied Sciences, Harvard University, Cambridge, MA 02138, USA.}

\address[1]{Laboratory of Machine Learning and Intelligent Instrumentation, Federal University of Rio Grande do Norte, Natal 59078-970, Brazil.}
\address[2]{Federal Institute of Education, Science and Technology of Rio Grande do Norte, Paraiso, Santa Cruz, RN, 59200-000, Brazil.}
\address[3]{Department of Statistics and Operations Research, University of Granada, Spain.}
\address[4]{Department of Computer Engineering and Automation, Federal University of Rio Grande do Norte, Natal, RN, 59078-970, Brazil.}

\begin{abstract}
The amount of data in real-time, such as time series and streaming data, available today continues to grow. Being able to analyze this data the moment it arrives can bring an immense added value. However, it also requires a lot of computational effort and new acceleration techniques. As a possible solution to this problem, this paper proposes a hardware architecture for Typicality and Eccentricity Data Analytic (TEDA) algorithm implemented on Field Programmable Gate Arrays (FPGA) for use in data streaming anomaly detection.  TEDA is based on a new approach to outlier detection in the data stream context. In order to validate the proposals, results of the occupation and throughput of the proposed hardware are presented. Besides, the bit accurate simulation results are also presented. The project aims to Xilinx Virtex-6 xc6vlx240t-1ff1156 as the target FPGA.
\end{abstract}


\end{frontmatter}


\section{Introduction}

Outlier detection or anomaly detection consists in detect rare events in a data set. It is a central problem in many application areas such as time series forecasting, data mining and industrial process monitoring. Due the increasing number of sensors in the most diverse areas and applications, there is a huge raise in the availability of data from time series. Thus, outlier detection for temporal data has become a central problem \cite{outlier_detection}, especially when data are captured and processed continuously in online way. In this case, the data are considered as data streams \cite{HTM}.

Some important aspects need to be considered when choosing an anomaly detection method, such as the computational effort to handle large streaming data. Since the received information need to be stored and analyzed without compromising memory and run-time. Many of the solutions presented in the literature require prior knowledge of the process and system, such as mathematical models, data distribution, and predefined parameters \cite{clauber}. Anomaly detection is traditionally done from statistical analysis, using probability and making a series of initial assumptions that in most cases are not in practice applied.

A disadvantage of the traditional statistical method is comparing a single point with the average of all points rather than comparing with sample or data pairs. This way, the information is no longer punctual and local. Moreover, probability theory was developed from examples where processes and variables are purely random. However, real processes are not purely random and shows dependency between samples. Thus, real problems are addressed from offline processes, where the entire data set needs to be known. Being a potential problem of the traditional method. Another problem with traditional approaches is that they often use an offline dataset. Thus, all samples must be previously available from the beginning of the algorithm execution \cite{Unsupervised}, making it impossible to use in real-time and data stream applications. This type of data presents new technical challenges and opportunities in new fields of work. Detecting real-time anomalies can provide valuable information in critical scenarios, but it is a high computational demand problem that still lacks reliable solutions capable of providing high processing capabilities.

Typicality and Eccentricity Data Analytic (TEDA) is based on new approach to outlier detection in data stream context \cite{typicality} and it can applied  with an algorithm to detect autonomous behavior in  industrial process operation, for example. TEDA analyzes the density of each sample of data read, calculated according to the distance from the sample to the other samples previously read. It is an online algorithm that learns autonomously without the need for prior knowledge about the process or parameters. Therefore, the computational effort required is smaller, allowing the use in real time applications \cite{clauber}.

TEDA can be used as an alternative statistical framework for analyzing most data, except for purely random processes. It is based on new metrics, all based on similarity/proximity of data in the data space, not in density or entropy, as in traditional methods. The metrics used with TEDA are typicality, defined in \cite{typicality} as the extent to which objects are ?good examples? of a concept, and eccentricity, defined as how distinct the object is from the rest of the group. A high eccentricity data has a low typicality and is usually an outlier \cite{clauber}.

Eccentricity can be very useful for anomaly detection, image processing, fault detection, particle physics, etc. Allows analysis for data samples (which can also be done in real time for data stream) \cite{anomaly}. It is also relevant in clustering processes, since elements of a cluster are naturally opposed to the atypical \cite{typicality}.

Another area where anomaly detection has been increasingly used is in industry 4.0 projects. One of the challenges of the Industry 4.0  is the detection of production failures and defects \cite{anomaly2}. New technologies aim to add value and increase process productivity, but face difficulties in performing complex and massive-scale computing due to the large amount of data generated \cite{bigdata1}. The huge accumulation of real time data to flow in a network, for example, can quickly overload traditional computing systems due to the large amount of data that originates from the sensors and the requirement for intensive processing and high performance. The development of specialized hardware presents itself as a possible solution to overcome the bottlenecks, making it possible to create solutions for mass data processing and, at the same time, consider ultra-low-latency, low-power, high-throughput, security and ultra-reliable conditions, important requirements for increasing productivity and quality in industry 4.0 processes.

Thinking about the challenges presented, this work proposes a specialized hardware architecture of TEDA for anomaly detection. The development of the hardware technique allows systems to be made even faster than their software counterparts, extending the possibilities of use for situations where time constraints are even more severe. In addition allowing its use in applications with large data processing. The works \cite{ssae,luciieee2019,torquato2019high,lopes2019parallel, Swarm} were developed in hardware, specifically on FPGA, for the acceleration of complex algorithms. The development of machine learning algorithms in hardware has grown significantly. This is justified from performance data with respect to system sampling times compared to software equivalents. One of the motivations for this work is the possibility of accelerating the TEDA algorithm and handling large data streams, such as streaming and real-time.

In this work, all validation and synthesis results was made using a FPGA Virtex 6 xc6vlx240t1ff1156. The FPGA choice was because it has high performance. Modern FPGAs can deliver performance and density comparable to Application Specific Integrated Circuits (ASICs), without the disadvantages of high development time and enabling reprogramming, as FPGAs have a flexible architecture. 

The rest of this paper is organized as follows: This first section presented a introduction about the work explaining the motivation behind it and major contributions. Section \ref{Sec:TEDArelatedwork} discusses some related works and the state of the art. In Section \ref{sec:teda} will be presented a theoretical foundation regarding the TEDA technique. Section \ref{Sec:TEDAimplementation} presents the implementation description details for the architecture proposed. Section \ref{Sec:TEDAresults} will present the validation and synthesis results of the proposed hardware, as well as comparisons with software implementations. Finally, Section \ref{Sec:TEDAconclusion} will present the conclusions regarding the obtained results.

\section{Related work}\label{Sec:TEDArelatedwork}

Real-time anomaly detection in data stream has potential applications in many areas. Such as: preventive maintenance, fault detection, fraud detection, signals monitoring, among others. Concepts that can be used in many different ranges of industry, such as information technology, finance, medicine, security, energy, e-commerce, agriculture, social media, among others. In the literature there are some uses of the TEDA technique for anomaly detection and even for classification.

The article presented in \cite{anomaly} shows a proposal for a new TEDA-based anomaly detection algorithm. The proposed method, called by the author $\sigma$ gap, combines the accumulated proximity information for all samples with the comparison of specific point pairs suspected of being anomalies. Using local spatial distribution information about the vicinity of the suspect point. In the journal, TEDA is compared to an approach using traditional statistical methods, emphasizing that the set of initial assumptions is different. TEDA has been shown to be a generalization of traditional statistics compared to a known analysis, n $\sigma$, which is a widely used principle for threshold anomaly detection. The same result was obtained for both approaches, although TEDA does not need the initial assumptions. In addition, for various types of proximity measurements (such as Euclidean, Cosine, Mahalanobis), it has been shown that due to the recursion feature, TEDA is computationally more efficient and suitable for online and real-time applications.

In the work \cite{Online} a study is presented about the use of TEDA for fault detection in industrial processes. The work is pioneering the use of this approach for real industry data. For the experiments, TEDA was applied online to the dataset provided by the DAMADICS (Development and Application of Methods for Actuator Diagnosis in Industrial Control Systems) database, one of the most widely used benchmarks in fault detection and diagnosis applications. The experiments showed good results both in accuracy and execution time, which shows the suitability of this approach for real-time industrial applications. Finally, it was found that the TEDA algorithm is capable of dealing with the limitations and particularities of the industrial environment.

The paper of \cite{Evolving} is intended to enable the use of TEDAClass, which consists of the TEDA algorithm for classification, in big data processing. The main feature of the proposed algorithm, called TEDAClassBDp, is the processing of block data, where each block uses the TEDAClass so that all blocks operate in parallel. As with TEDAClass, the proposed algorithm does not require information from previous data, and its operation occurs recursively, online and in real-time. The results indicated a reduction in time and computational complexity, without significantly compromising its accuracy, which indicates the strong possibility of using the proposal in problems where it is necessary to process large volumes of data quickly.

The work presented in \cite{tdfangelov} proposes a new non-frequency and density-based data analysis tool. Classified by the author as a further development of TEDA and an effective alternative to the probability distribution function (pdf). Typicality Distribution Function (TDF) can provide valuable information for extreme process analysis, fault detection and identification, where the number of extreme event or fault observations is often disproportionately small. The proposed offers a closed non-parametric analytical (quadratic) description, extracted from the actual realizations of the data exactly in contrast to the usual practice in which these distributions are being assumed or approximated. In addition, for various types of proximity and similarity measures (such as Euclidean, Mahalonobis, and cosine distances), it can be recursively calculated, thus computationally efficient and suitable for online and real-time algorithms. As a fundamental theoretical innovation, TDF and TEDA application areas can range from anomaly detection, grouping, classification, prediction, control, filter regression (similar to Kalman). Practical applications may be even broader, so it is difficult to list them all.

The paper presented in \cite{clauber} proposes the application of TEDA for fault detection in industrial processes. The effectiveness of the proposal has been demonstrated with two real industrial plants, using data streaming, and compared with traditional failure detection methods. This paper presents a practical application of the TEDA algorithm for two fault detection problems of real industrial plants. The first application uses a well-known database, DAMADICS, a database that provides actual data on the water evaporation process of an operating plant of a Polish sugar manufacturing plant. The second application was made from data analysis of a pilot plant of the authors' university laboratory. A plant equipped with real industrial instruments used for process control.

The work of \cite{Unsupervised} presents a new proposal for unsupervised fuzzy classifier, capable of aggregating the main characteristics of evolving classifiers, as well as making fuzzy classifications of real time data streams completely online. The proposed algorithm uses TEDA concepts, replacing traditional clusters with data clouds, granular structures without shape or predefined boundaries. For data classification, the proposed approach uses the concepts of soft-labeling rather than mutually exclusive classes. Experiments performed using data obtained from different operational failures of a real industrial plant, showed very significant results regarding unsupervised as well as semi-supervised learning, requiring minimal human interaction.

The manuscript presented in \cite{HTM} brings a new algorithm for detecting anomalies based on an online memory sequence algorithm called \textit{Hierarchical Temporal Memory (HTM)}. The performance of the proposed algorithm was evaluated and compared with a set of real time anomaly detection algorithms. Comparative analysis was performed as a way to evaluate anomaly detection algorithms for data streaming. All analyzes were performed from the \textit{Numenta Anomaly Benchmark (NAB)} \cite{numenta}, which is a benchmark of actual streaming data.

The paper published by \cite{tedanetwork} brings a study for anomaly detection in TCP / IP networks. The purpose of the paper is to detect computer network anomalies in the process of virtual machine (VM) live migration from local to cloud, by comparing this approach between TEDA, clustering K-Means, and static analysis. They used the tuple - \textit{Source IP, Destination IP, Source Port, and Destination Port} - to create a signature process and validate errors, including those of traffic flow hidden in the legitimate network. Testing was done using the SECCRIT (SEcure Cloud Computing for CRitical Infrastructure IT - http://www.seccrit.eu) project dataset, which allows anomalies or environmental attacks to be analyzed with Live Migration and other background traffic conditions.
The results demonstrate that the proposed method makes it possible to automatically and successfully detect anomalies in attacks, network port scan (NPS) and network scan (NS).
A major difficulty is distinguishing a high-volume attack from a denial of service (DoS) attack, for example. Accuracy and false negative rate calculations were made for comparison with K-Means and the proposed solution, with TEDA having better rates in almost all measurements performed.

As the amount of data that needs to be processed grows exponentially and autonomous systems become increasingly important and necessary. Implementation of machine learning and streaming algorithms have been studying in literature. The work presented in \cite{adaptive} describes how to use run-time reconfiguration on FPGAs to improve the efficiency of streaming data transmission in shared communication channel with real-time applications. The reconfigurable architecture proposed consists of two subsystems: the reconfiguration subsystem, which running the modules, and the scheduling subsystem, that controls which modules are loaded to the reconfiguration subsystem.

Besides, many works in the literature have been studied fault and anomaly detection in hardware. In work \cite{TargetAnomalyDetection}, an implementation of target and anomaly detection algorithms for real-time hyper-spectral imaging was proposed on FPGA. The algorithms were implemented in streaming fashion, similar to this work. The results, obtained from a Kintex-7 FPGA using fixed point structure, were very satisfactory and demonstrated that the implementation can be used in different detection circumstances. The work \cite{ECGAnomalyDetection} presented a study of the impact of Neural Network architectures compared to statistical methods in the implementation of an Electrocardiogram (ECG) anomaly detection algorithm on FPGA. The fixed point implementation contributes to reduce the amount of needed resources. However, the design was made with High Level Sinthesys (HLS), witch could not optimize the FPGA resources consumption. In relation to the TEDA algorithm, no studies in the literature aimed at exploring its hardware implementation on FPGA were identified  to date this paper had been write, which this work proposes to accomplish in a pioneering manner.

\section{TEDA}\label{sec:teda}

TEDA was introduced by \cite{angelov2014outside} as a statistical framework, influenced by recursive density estimation algorithms. However, unlike algorithms that uses data density as a measure of similarity, TEDA uses concepts of typicity and eccentricity to infer whether a given sample is normal or abnormal to the dataset. The methodology used in TEDA does not require the use of a previous data information, and can be applied to problems involving fault detection, clustering, classification, among others \cite{angelov2014outside}.

TEDA is a data structure-based anomaly detection algorithm that aims to generalize and avoid the need for well-known, but very restrictive, initial conditions inherent in traditional statistics and probability theory \cite{outside}. The approach presented in the TEDA has some advantages over traditional statistical anomaly detection methods. Its recursive feature allows it to handle large volumes of data, such as data streams, with low computational cost and online, enabling faster processing.

TEDA main features include \cite{anomaly}:
\begin{itemize}
\item It is entirely based on data and its distribution in data spaces;
\item No previous assumptions are made;
\item Limits and parameters does not need to be pre-specified;
\item No sample independence required;
\item An infinite number of observations are not required.
\end{itemize}

The typicality of TEDA is the similarity of a given data sample to the rest of the dataset samples to which it belongs. Eccentricity, on the other hand, is the opposite of typicality, which indicates how much a sample is dissociated from the other samples in its set. Thus, an outlier can be defined as a sample with high eccentricity and low typicality, considering a threshold established for comparison. It is important to note that for eccentricity and typicality calculations no parameter or threshold is required.

To calculate the eccentricity of each sample, TEDA uses the sum of the geometric distances between the analyzed sample $\bm{x}_k$ and the other samples in the set. Thus, the higher this value, the greater the eccentricity of the sample, and consequently, the lower its typicality. \cite{anomaly} proposed recursively calculating eccentricity. Thus, the eccentricity, $\xi$ can be expressed as
\begin{equation}
\label{excent}
\xi_{k} (x) = \frac{1}{k} + \frac{( \bm{\mu} _{k}^{x}-\bm{x}_k)^T(\bm{\mu}^{x}_{k}-\bm{x}_k)}{k[\sigma^2]^{x}_{k}}, [\sigma^2]^{x}_{k}>0  
\end{equation}
where $k$ is discreization instant; $\bm{x}_{k}$ is a input set of N elements in the k-th iteration, $\bm{x}_{k}=[x_k^1 \ x_k^2 \ ... \ x_k^N]$; $\bm{\mu}_{x}^{k}$ is also a N elements vector, equal to the average of $\bm{x}_k$ at the $k$-th iteration and $[\sigma^2]^{x}_{k}$ is the variance of $\bm{x}_k$ at the $k$-th iteration. The calculation of $\bm{\mu}_{x}^{k}$ and $[\sigma^2]^{x}_{k}$ is also recursively done, using the following equation
\begin{equation}
\label{media}
\bm{\mu} ^{x}_{k} = \frac{(k-1)}{k}\bm{\mu} ^{x}_{k-1}+\frac{1}{k}\bm{x}_{k},\ k \geq 1,\ \bm{\mu} ^{x}_{0} = 0
\end{equation}
and 
\begin{equation}
\label{sigma}
[\sigma^2]^{x}_{k} = \frac{(k-1)}{k}[\sigma^2]^{x}_{k-1}+\frac{1}{k}\left \| \bm{x}_{k}-\bm{\mu} _k \right \|^2,\ k \geq 1,\ [\sigma^2]^{x}_{0} = 0.
\end{equation}

The typicality of a given sample $\bm{x}_k$, at the $k$-th iteration, can be expressed as a complement to eccentricity \cite{anomaly}, as follows 
\begin{equation}
\label{tipicidade}
\tau _k(x) =1-\xi _k(x).
\end{equation}

In addition, \cite{anomaly} also defined that normalized eccentricity can be calculated as 
\begin{equation}
\label{excent_normal}
\zeta_{k} (x) = \frac{\xi_{k} (x)}{2}, \sum^{k}_{i=1} \xi_{k} (x)=1,\ k \geq 2. 
\end{equation}

In order to separate normal state data from abnormal state data, it is necessary to define a comparison threshold. For anomaly detection, the use of the $m\sigma$ \cite{mo} threshold is widespread. However, this principle must first assume the distribution of the analyzed data, such as the Gaussian distribution \cite {anomaly}. Chebyshev inequality can be used for any data distribution, assuming that the probability that the data samples are more than $m\sigma$ from the average is less than or equal to $1/m^2$, where $\sigma$ is the standard deviation of the data \cite{saw1984chebyshev}.

The condition that produces the same results as Chebyshev's inequality, discarding any assumptions about data and its independence, can be expressed as \cite{anomaly} 
\begin{equation}
\label{excent_limiar}
\zeta_{k} > \frac{m^2 + 1}{2k},\ m > 0
\end{equation}
where $m$ corresponds to the comparison threshold.

For a better understanding of the hardware implemented technique in this work, the Algorithm \ref{alg_teda} details the operation of TEDA, based on the equations presented above.

\begin{algorithm}[h]
\scriptsize
\LinesNumbered
\SetNlSty{}{}{:}
\SetAlCapFnt{}
\footnotesize
\caption{TEDA}
\label{alg_teda} \BlankLine
\KwIn{$\mathbf{x}_{k}$: $k$-th sample; $m$: threshold} \label{alg_input}
\KwOut{outlier: sample classification as abnormal or normal} \BlankLine
\Begin{ \BlankLine
\While{receive $\mathbf{x}$}
{
	\If{k=1}{ \label{alg:primeiraamostra}
		$\bm{\mu} ^{x}_{k} \leftarrow \mathbf{x}_{k}$\;
		$[\sigma^2]^{x}_{k} \leftarrow 0$\; \label{alg:primeiraamostrasigma}
	}
	\Else{
	    update $\bm{\mu} ^{x}_{k}$ using equation \ref{media}\; \label{alg_media}
	    update $[\sigma^2]_{k}^{x}$ using equation \ref{sigma}\; \label{alg_sigma}
	    update $\xi_{k}(x)$ using equation \ref{excent}\; \label{alg_excent}
	    update $\zeta_{k}(x)$ using equation \ref{excent_normal}\; \label{alg_excent_normal}
		\If{$\zeta_{k}(x) > \frac{m^2+1}{2k}$}{ \label{alg_outlier}
    	    $outlier \leftarrow true$\; \label{alg_outlier2}
    	}
		\Else{		\label{alg_outlier3}                        
		    $outlier \leftarrow false$\; \label{alg_outlier4}
		}
	}
	$k \leftarrow k+1$\;
}
} \BlankLine
\end{algorithm}

As presented in the Algorithm \ref{alg_teda}, only input data samples, $\bm{x}_k$, and a comparison threshold, $m$, are used as input to the algorithm. The output for each entry, $\bm{x}_{k}$, is the indication of the sample's classification as abnormal (outlier = true) or normal (outlier = false).

\section{Implementation description} \label{Sec:TEDAimplementation}

In this work, a TEDA FPGA proposal was implemented using Register Transfer Level (RTL) such as works presented in \cite{ssae,luciieee2019,torquato2019high,lopes2019parallel, Swarm}. In the following section characteristics of the proposal will be presented, as well as details regarding processing time.  A design overview can be seen in Figure \ref{GeralTEDA}. 

\subsection{Architecture proposal overview} \label{sub:TEDAgeral}

\begin{figure}[ht]
\begin{center}
  \includegraphics[width=1\linewidth]{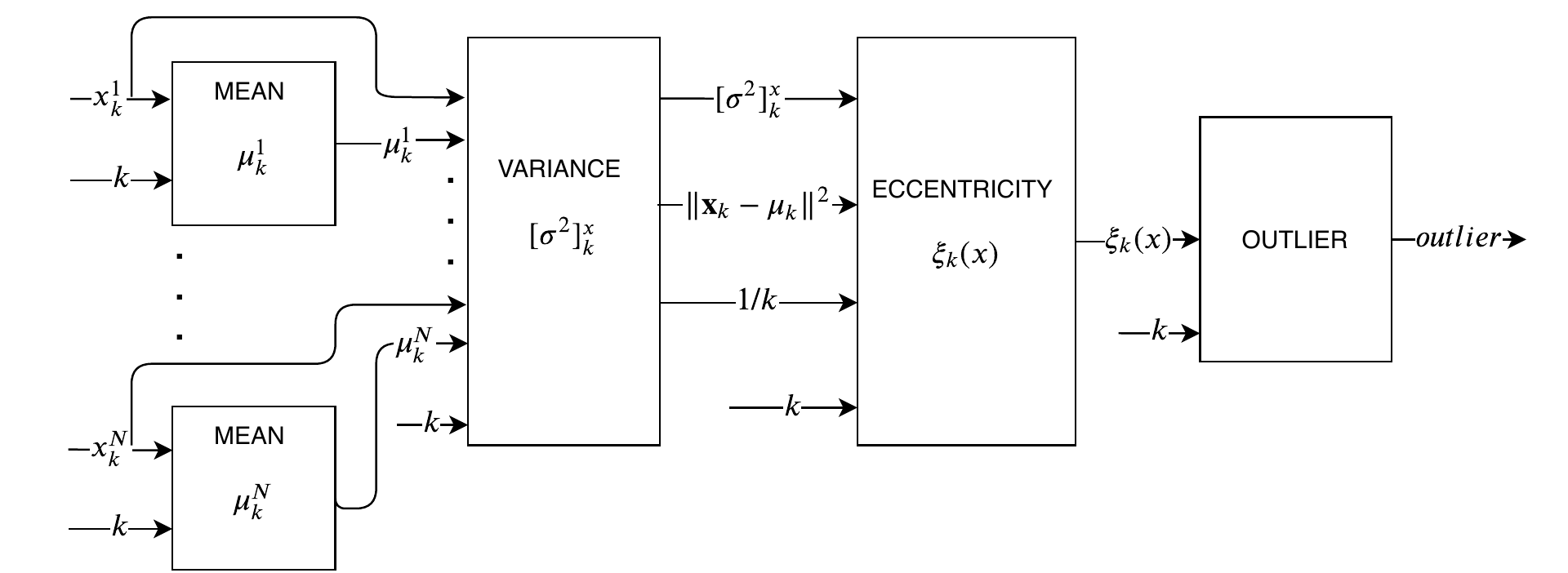}\\
  \caption{General architecture overview.}\label{GeralTEDA}
\end{center}
\end{figure}

As illustrated in the Figure \ref{GeralTEDA}, the proposed implementation of TEDA has four different block structures: The MEAN module, which implements the average described in Equation \ref{media}; The VARIANCE module, responsible for calculate the variance as presented at the equation \ref{sigma}; The ECCENTRICITY module, which calculates the eccentricity, as presented in the equation \ref{excent}; and the OUTLIER module, a block used to normalize the eccentricity as in equation \ref{excent_normal} and compare with the threshold, as showed in equation \ref{excent_limiar}. The architecture was developed in an attempt to pipeline the operations presented in Algorithm \ref{alg_teda} in order to decrease the TEDA processing time. So, the output of the ECCENTRICITY and OUTLIER modules are one clock cycle delayed in relation to VARIANCE module and two in relation to MEAN module. As well as VARIANCE module is one clock cycle delayed in relation to MEAN module. Each of the modules are detailed later in the following sections.

The implementation has the Algorithm \ref{alg_teda} as reference. The system receives the FPGA clock and the $k$-th sample vector $\mathbf{x}_{k}$ as inputs. The $k$-th iteration number is updated from the increment of a counter and the $m$ threshold is used as a constant, stored at OUTLIER module. As in the algorithm  line \ref{alg_media}, the MEAN module do each single element average of $\mathbf{x}_{k}$ vector. It is possible to observe that there are $N$ MEAN blocks, where $N$ is the vector size. This block will be detailed in section \ref{sub:mean}. After this step, moving to the next line (\ref{alg_sigma}), the calculation of variance is done in VARIANCE Module, this block is detailed in the section \ref{sub:variance}. ECCENTRICITY block has as inputs the signals that left the block VARIANCE and $k$, as referred at line \ref{alg_excent} and detailed in subsection \ref{sub:excent}. OUTLIER block is detailed in subsection \ref{sub:outlier}. It receives the ECCENTRICITY, $\xi _k(x)$, and calculate the normalized eccentricity to compare with the threshold as presented in lines \ref{alg_excent_normal}, \ref{alg_outlier} and \ref{alg_outlier2}. 

\subsection{Module I - MEAN}
\label{sub:mean}

\begin{figure}[ht]
\begin{center}
  \includegraphics[width=1\linewidth, trim={1cm 0 1cm 0},clip]{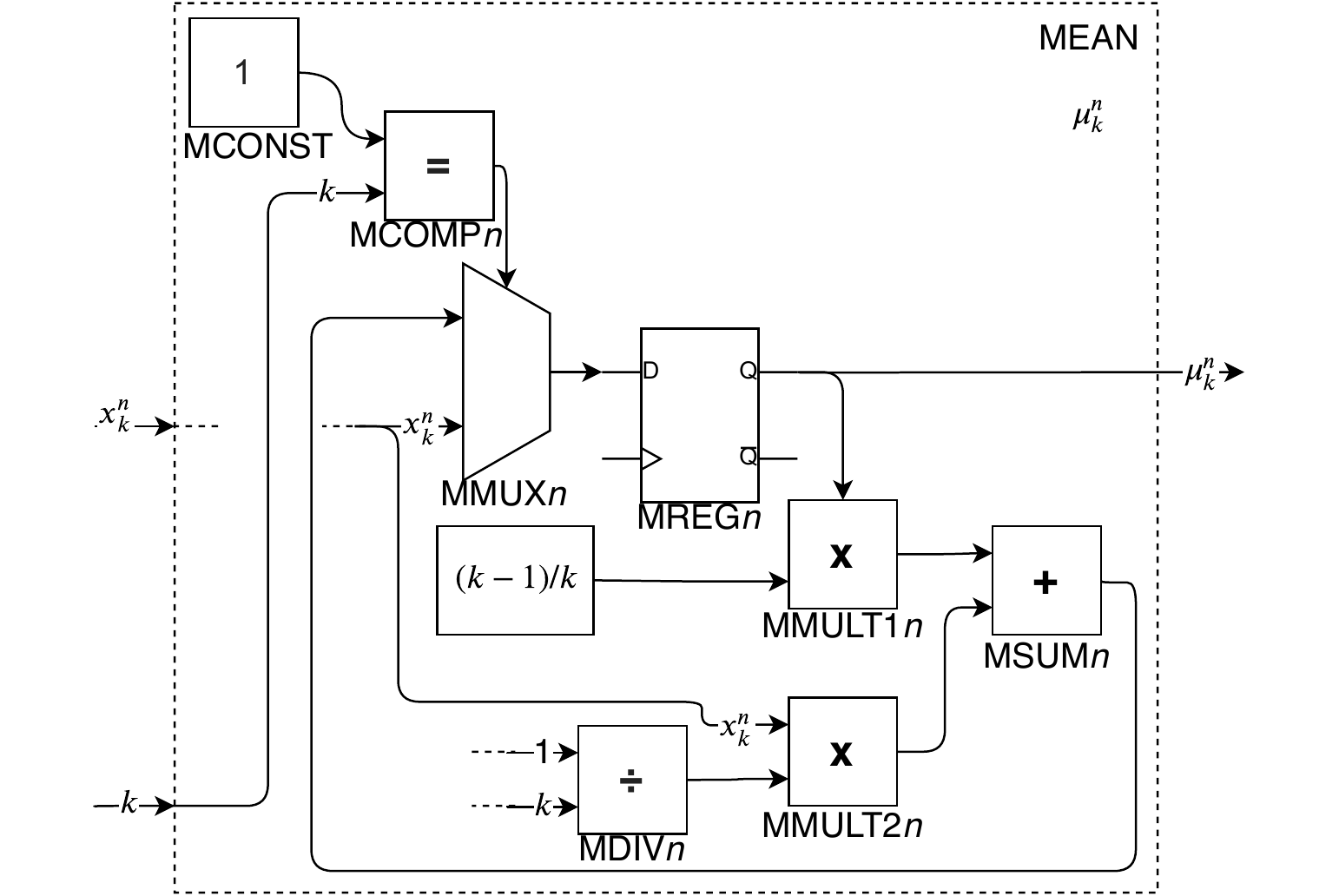}\\
  \caption{MEAN module.}\label{MeanTEDA}
\end{center}
\end{figure}

Each \textit{n}-th MEAN module computes the average of each one of n-th elements vector $\bm{x}_{k}$ acquired at run time. The implementations is based on Equation \ref{media} and it is detailed in Figure \ref{MeanTEDA}.  In addition to receiving the \textit{n}-th element of vector $\bm{x}_{k}$ as an input, the MEAN block uses a counter to define the number of sample interaction, $\textit{k}$. The implementation uses a comparator block identified at the Figure \ref{MeanTEDA} as MCOMP\textit{n} witch is used to verify if the system is in the first iteration as  Line \ref{alg:primeiraamostra} of Algorithm \ref{alg_teda}. The MMUX\textit{n} is a multiplexer that acts as a conditional evaluation, using as selecting value the output of MCOMP\textit{n} comparator.  The register MREG\textit{n} is storing the \textit{n}-th $\bm{\mu} ^{x}_{k}$ element ($\mu ^{n}_{k}$). The $\mu ^{n}_{k}$ value stored in MREG\textit{n} is multiplied with $\frac{k-1}{k}$ in MMULT1\textit{n} and added in MSUM\textit{n} with the output of MMULT2\textit{n} that has as input  $x ^{n}_{k}$ and the inverse value of $k$. Each n-th element of vector $\bm{x}_{k}$, $x ^{n}_{k}$,  requires a MEAN block.
 
\subsection{Module II - VARIANCE}
\label{sub:variance}

\begin{figure}[ht]
\begin{center}
  \includegraphics[width=1\linewidth]{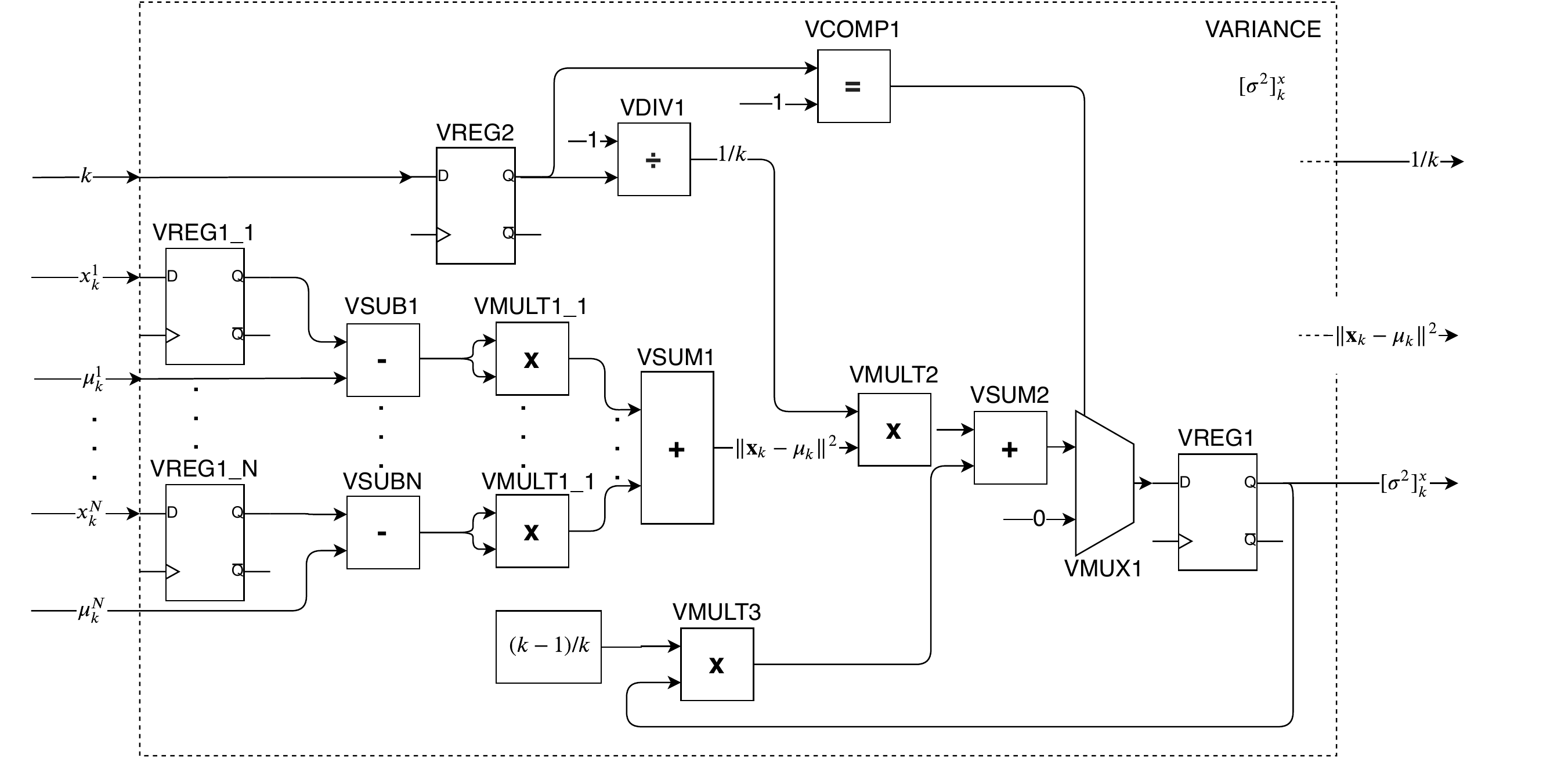}\\
  \caption{VARIANCE module.}\label{VarianceTEDA}
\end{center}
\end{figure}

The VARIANCE module is illustrated in Figure \ref{VarianceTEDA}. It computes the variance of $\bm{x}_{k}$ vector samples by receiving the $\bm{x}_{k}$ vector itself and its average, $\bm{\mu} ^{x}_{k}$, calculated in the previous MEAN blocks. 

The VARIANCE module, as the MEAN module, uses a comparator identified at the Figure \ref{VarianceTEDA} as VCOMP1 also to verify if the system is in the first iteration (Line \ref{alg:primeiraamostra} of Algorithm \ref{alg_teda}). The VMUX1 is a multiplexer that also implements a conditional evaluation to release the value $0$ in the register output VREG1 in the first iteration. The register VREG1 stores the variance value, $[\sigma^2]^{x}_{k}$, from the second iteration. The other registers in the block, VREG2 register and the N VREG$n$ registers, are used to delay by one clock cycle the iteration number $k$ and the elements of $\bm{x}_{k}$ respectively.

As demonstrated in Equation \ref{VarianceTEDA}, the variance calculation is done recursively. It is necessary to calculate  $\left \| \bm{x}_{k}-\bm{\mu} _{k} \right \|^2$ and to do that, N subtractors (VSUB$n$) and N multipliers (VMULT1\_$n$) are used, as well a adder (VSUM1) with N inputs. Each element of vector $\bm{\mu} ^{x}_{k}$ is subtracted from its respective element in vector $\bm{x}_{k}$ and the result of this operation is multiplied by itself (squared) and then added to the other results. The $\left \| \bm{x}_{k}-\bm{\mu} _{k} \right \|^2$ value is the multiplied (at VMULT2) by $1/k$. It is then added at VSUM2 adder with the variance calculated in the previous iteration, $[\sigma^2]^{x}_{k}$, multiplied (VMULT3) by $(k-1)/k$. From the second iteration on, this value passes through the VMUX1 multiplexer to the VREG1 register, delivering the calculation of the variance value at the VARIANCE block output. The values of $\left \| \bm{x}_{k}-\bm{\mu} _{k} \right \|^2$ and $1/k$ are also delivered at the output of the VARIANCE block to avoid redundant operations as they will be used in the next block, the ECCENTRICITY block. 

\subsection{Module III - ECCENTRICITY}
\label{sub:excent}

The ECCENTRICITY module is a simpler block than those previously presented. This is because it uses operations already performed in the VARIANCE block to calculate eccentricity. The geometric distance $\left \| \bm{x}_{k}-\bm{\mu} _{k} \right \|^2$ (equivalent to $(\bm{\mu} _{k}^{x}-\bm{x}_k)^T(\bm{\mu}^{x}_{k}-\bm{x}_k)$) is stored in register EREG3 and $1/k$ is stored in EREG4 register. As the ECCENTRICITY module is the architecture design of Equation \ref{excent} (Algorithm \ref{alg_teda} line \ref{alg_excent}), the variance value $[\sigma^2]^{x}_{k}$ is multiplied by $k$ (EMULT1) and  used to divise (EDIV1) the geometric distance $(\bm{\mu} _{k}^{x}-\bm{x}_k)^T(\bm{\mu}^{x}_{k}-\bm{x}_k)$. This operation output is added to $1/k$ in the ESUM1 adder, calculating the eccentricity of the samples  ($\xi_{k} (x)$) and delivering to the ECCENTRICITY block output.

\begin{figure}[ht]
\begin{center}
  \includegraphics[width=1\linewidth, trim={2cm 0 0 0},clip]{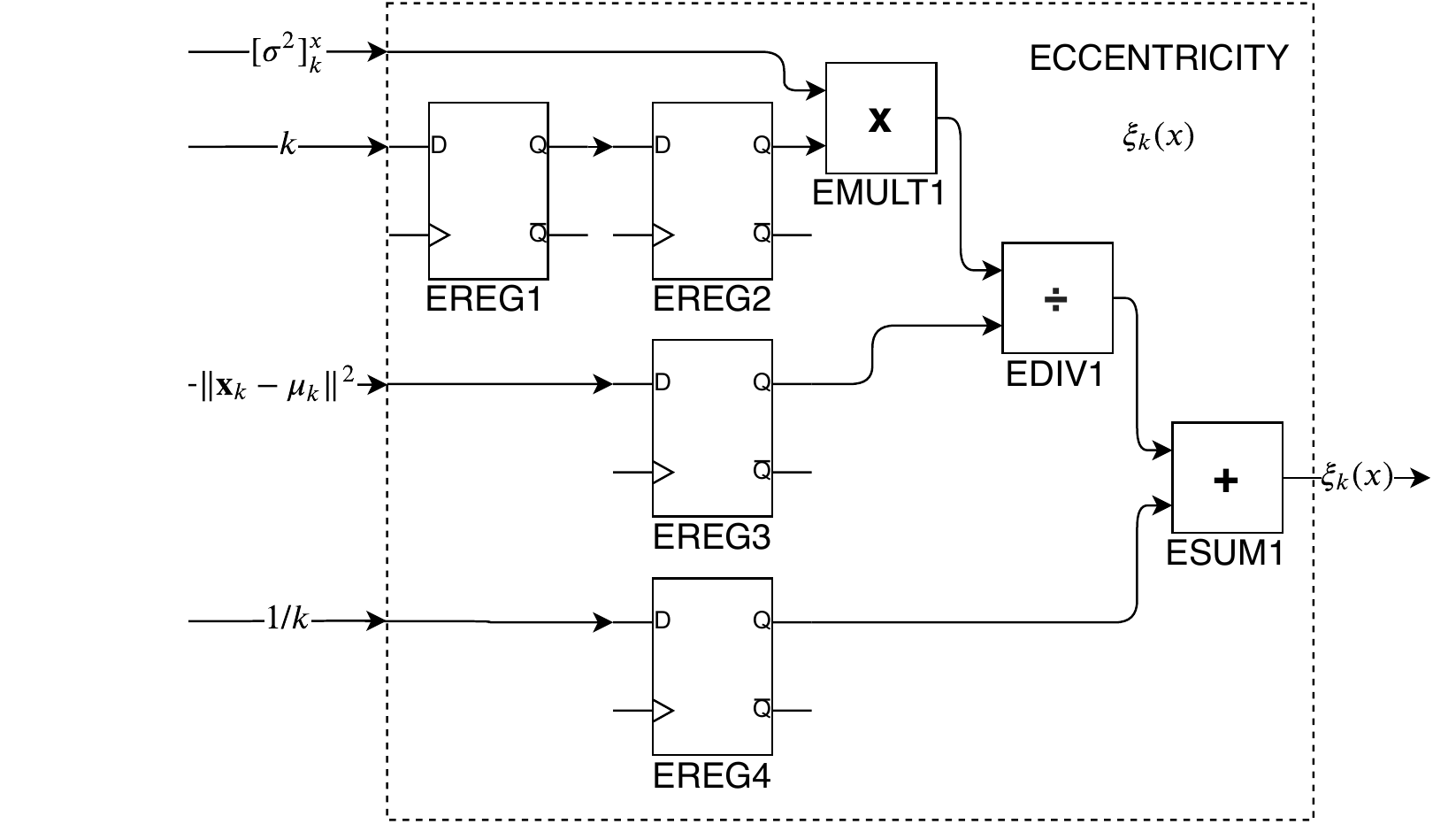}\\
  \caption{ECCENTRICITY module.}\label{ExcentricityTEDA}
\end{center}
\end{figure}

\subsection{Module IV - OUTLIER}
\label{sub:outlier}
\begin{figure}[ht]
\begin{center}
  \includegraphics[width=1\linewidth, trim={0.5cm 0 0.5cm 0},clip]{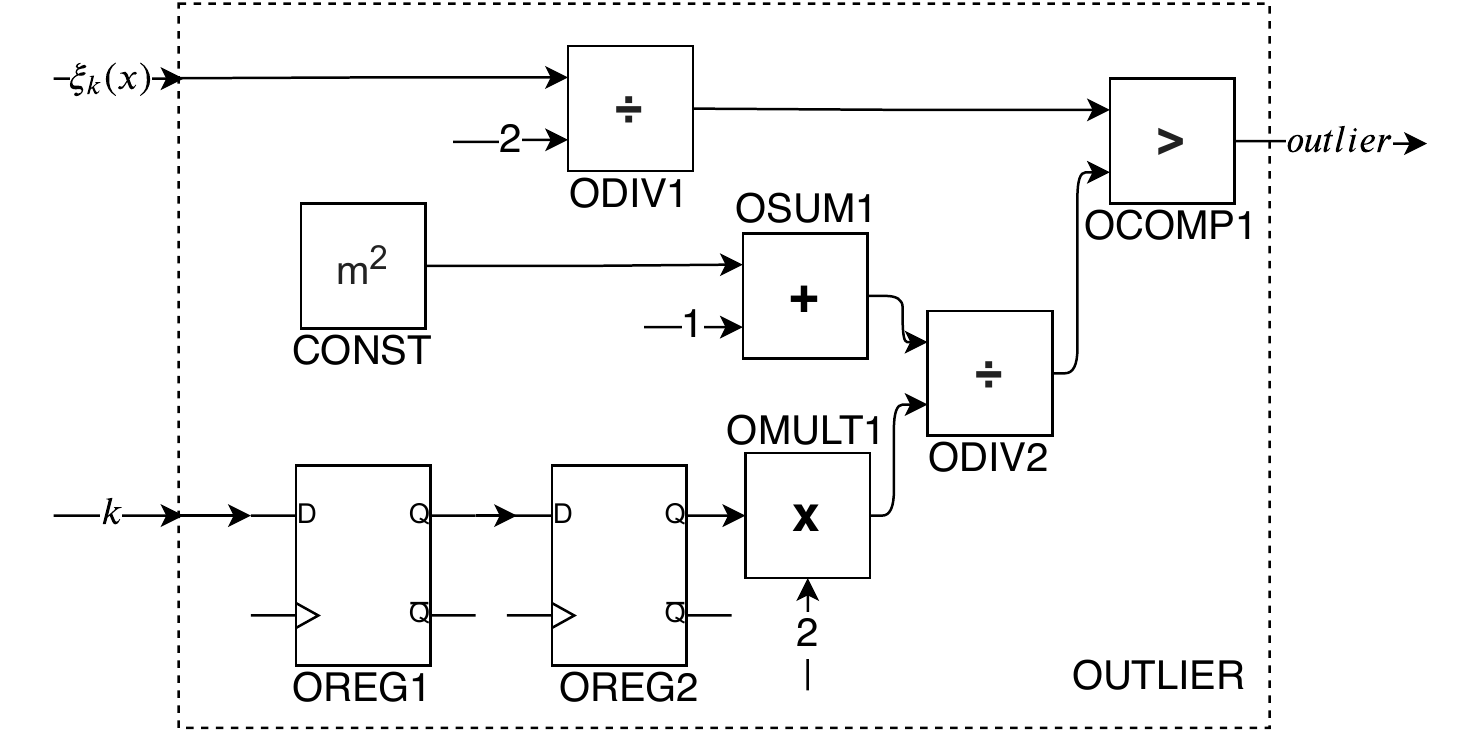}\\
  \caption{OUTLIER module.}\label{OutlierTEDA}
\end{center}
\end{figure}

Finally, in the OUTLIER block, the samples are classified into abnormal (outlier = true) or normal (outlier = false). The design module can be seen in Figure \ref{OutlierTEDA}. 
To classify the samples, the OUTLIER block normalizes eccentricity by dividing (ODIV1) by a constant, as shown in Equation \ref{excent_normal}, and compares (OCOMP1 this normalized eccentricity with a threshold as shown in the Lines \ref{alg_outlier}, \ref{alg_outlier2}, \ref{alg_outlier3} and \ref{alg_outlier4} of the Algorithm \ref{alg_teda}. The register OREG1 and OREG2 are used to synchronize the iteration number $k$, since as the modules act in pipeline, the operations carried out in the OUTLIER block (as well as in ECCENTRICITY module) are delayed by two clock cycles in relation to the system input.

\subsection{Processing time}

The proposed architecture has an initial delay, $d$, that can be expressed as
\begin{equation}
\label{delay}
d = 3 \times t_c 
\end{equation}
where $t_c$ is the system critical path time.

The execution time of the circuit implemented for TEDA algorithm is determined by the system critical path time, $t_c$. So, after the initial delay, the execution time of the proposed TEDA, $t_{TEDA}$, can be expressed as 
\begin{equation}
\label{tTEDA}
t_{TEDA} = t_c 
\end{equation}
thus, in every $t_{TEDA}$ it is possible to obtain the output of a sample inserted, that is, the sample classification as abnormal or normal.

The throughput of the implementation, $th_{TEDA}$, in samples per second (SPS) can be expressed as
\begin{equation}
\label{thTEDA}
th_{TEDA} = \frac{1}{t_{TEDA}} .
\end{equation}

\section{Results} \label{Sec:TEDAresults}

In this section will be presented the hardware validation and synthesis results for the architecture proposed in this work. All cases were validated and synthesized on floating point. Validation results were used to verify the hardware functionality, while synthesis results allow the system to be analyzed for important parameters for the design of hardware architectures such as hardware occupancy and processing time, considering factors such as throughput and speedup.

\subsection{Validation results} \label{TEDAsimulation}

To validate the hardware architecture of the TEDA algorithm, we used the DAMADICS (Development and Application of Methods of the Actuator Diagnosis in Industrial Control Systems) benchmark dataset \cite{damadics}. The benchmark provides a real data set of the water evaporation process in a Polish sugar factory. It is a plant with three actuators; a control valve, which controls the flow of water in the pipes; a pneumatic motor, which controls variable valve openings and a positioner. This dataset has faults at different times of the day on specific days. There are four different fault types, as shown in Table \ref{tab:FalhaTEDA}.

\begin{table}[ht]
\begin{center}
    \caption{Fault types \cite{damadics}.}\label{tab:FalhaTEDA}
\begin{tabular}{cc}
\hline
\textbf{Fault} & \textbf{Description} \\ 
\hline
f16 & Positioner supply pressure drop  \\ 
f17 & Unexpected pressure change across the valve \\
f18 & Fully or partly opened bypass valves  \\ 
f19 & Flow rate sensor fault \\ 
\hline
\end{tabular}
\end{center}
\end{table}

Artificial failures were introduced on specific days to plant operation data. The dataset has a set of $19$ faults in these $3$ actuators. As a way to validate the architecture, actuator $1$ failures were simulated. Table \ref{tab:iten} shows a detailed description of some introduced faults for actuator $1$.

\begin{table}[ht]
\begin{center}
    \caption{List of artificial failures introduced to actuator 1 \cite{damadics}.}\label{tab:iten}
\begin{tabular}{ccccp{4.3cm}}
\hline
\textbf{Item} & \textbf{Fault} & \textbf{Sample} & \textbf{Date} & \textbf{Description} \\ 
\hline
1 & f18 &58800-59800 &Oct 30, 2001 &Partly opened bypass valve \\ 
2 & f16 &57275-57550 &Nov 9, 2001 &Positioner supply pressure drop \\ 
3 &f18 &58830-58930 &Nov 9, 2001 &Partly opened bypass valve  \\ 
4 &f18 &58520-58625 &Nov 9, 2001 &Partly opened bypass valve  \\ 
5 &f18 &54600-54700 &Nov 17, 2001 &Partly opened bypass valve  \\ 
6 &f16 &56670-56770 &Nov 17, 2001 &Positioner supply pressure drop  \\ 
7 &f17 &37780-38400 &Nov 20, 2001 &Unexpected pressure drop across the valve \\ 
\hline
\end{tabular}
\end{center}
\end{table}

\begin{figure}[ht]
\begin{subfigure}{.5\textwidth}
  \centering
  \includegraphics[width=1\linewidth]{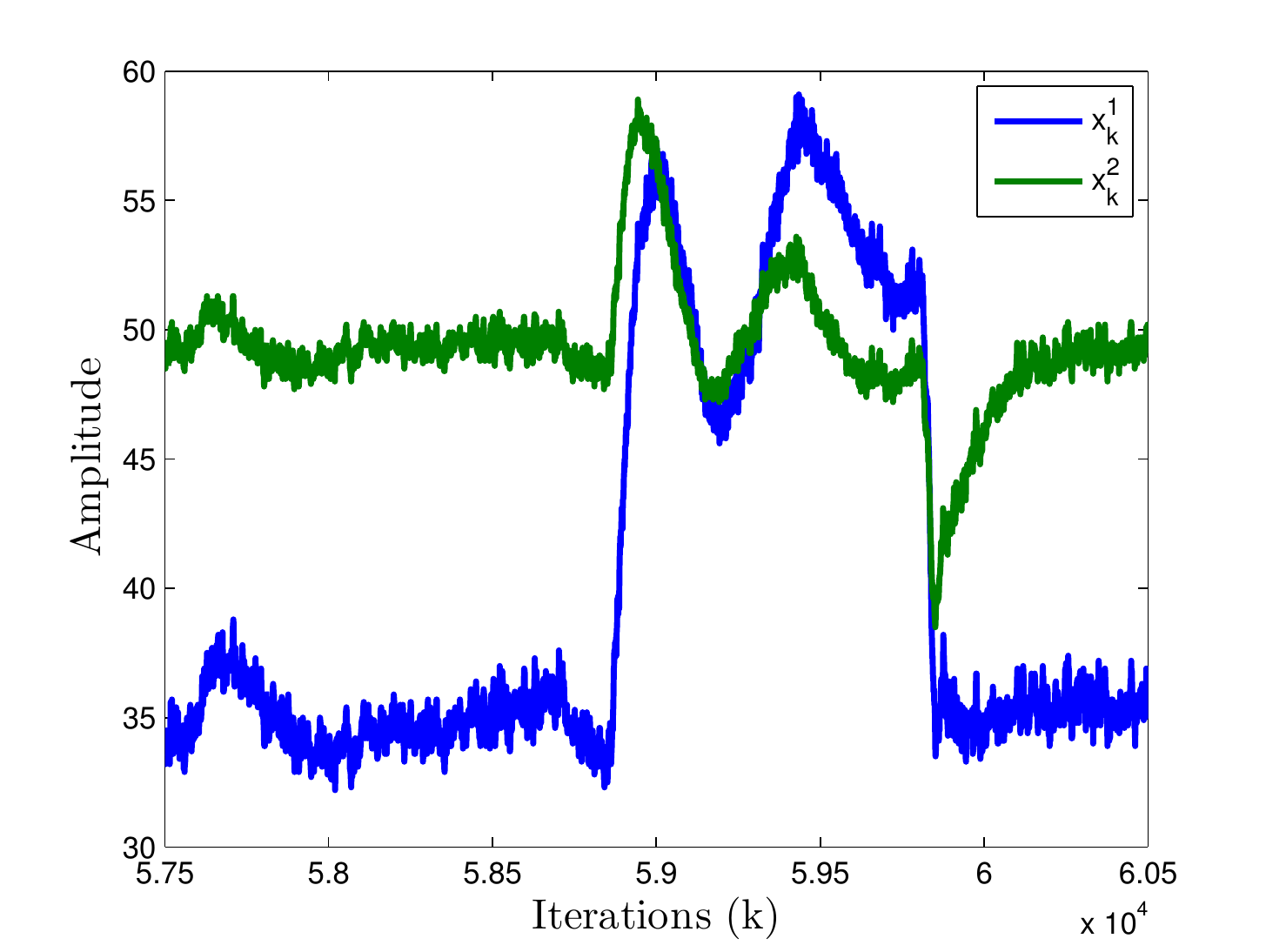}
   \caption{Fault item 1 - input vector $\bm{x}_{k}$.}\label{fig:Falha1_v}
\end{subfigure}
\begin{subfigure}{.5\textwidth}
  \centering
  \includegraphics[width=1\linewidth]{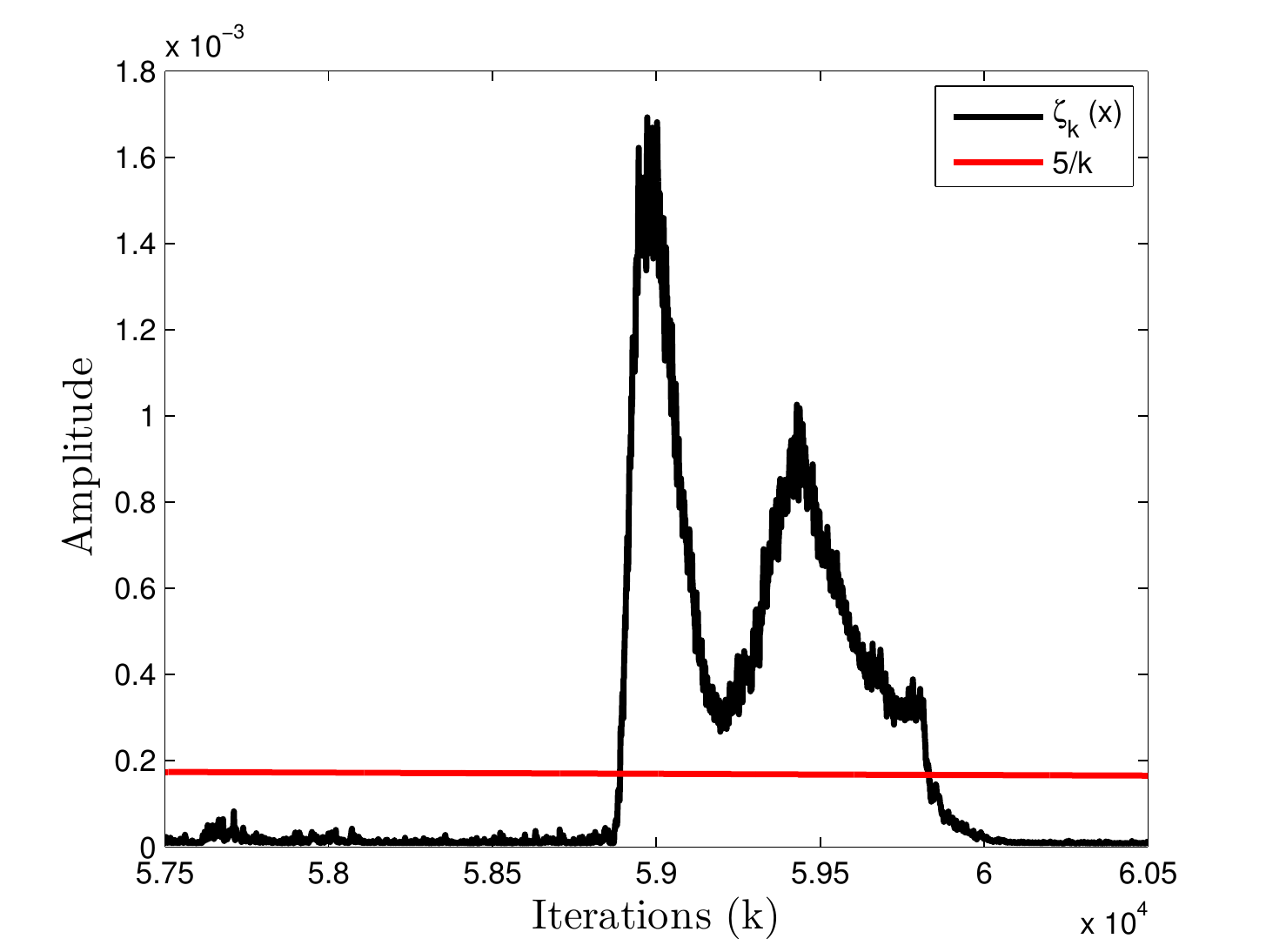}
   \caption{Fault item 1 - normalized eccentricity  $\zeta_{k} (x)$ with $5/k $ $(m=3)$ threshold.}\label{fig:Falha1_e}
\end{subfigure}
\caption{Detection of outliers in the dataset: Behavior of fault item 1.}\label{fig:Falha1}
\end{figure}

Figure \ref{fig:Falha1} shows the results obtained for the item 1 signal of Table \ref{tab:iten}. Figure \ref{fig:Falha1_v} illustrates the behavior of two simulated input variables in hardware architecture ($\bm{x}_{k}=[x_k^1 \ x_k^2]$). It is possible to observe that a failure happens between the moments $k$=58900 and $k$=59800. In Figure \ref{fig:Falha1_e} it is possible to observe that there is a sudden change in the behavior of the eccentricity (black curve), surpassing the value of the comparison threshold with $m=3$ (red curve).

\begin{figure}[ht]
\begin{subfigure}{.5\textwidth}
  \centering
  \includegraphics[width=1\linewidth]{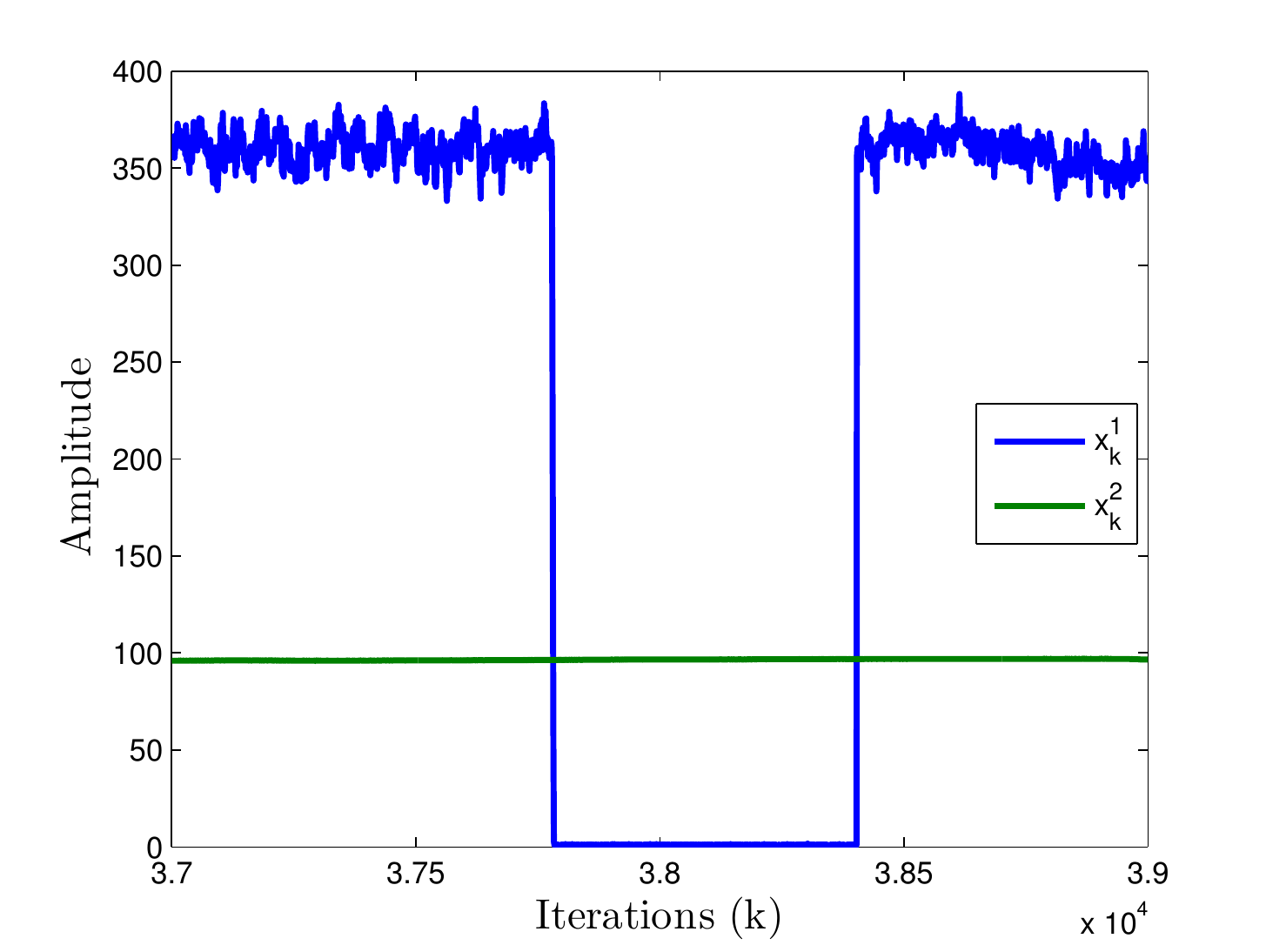}  
   \caption{Fault item 7 - input vector $\bm{x}_{k}$.}\label{fig:Falha7_v}
\end{subfigure}
\begin{subfigure}{.5\textwidth}
  \centering
  \includegraphics[width=1\linewidth]{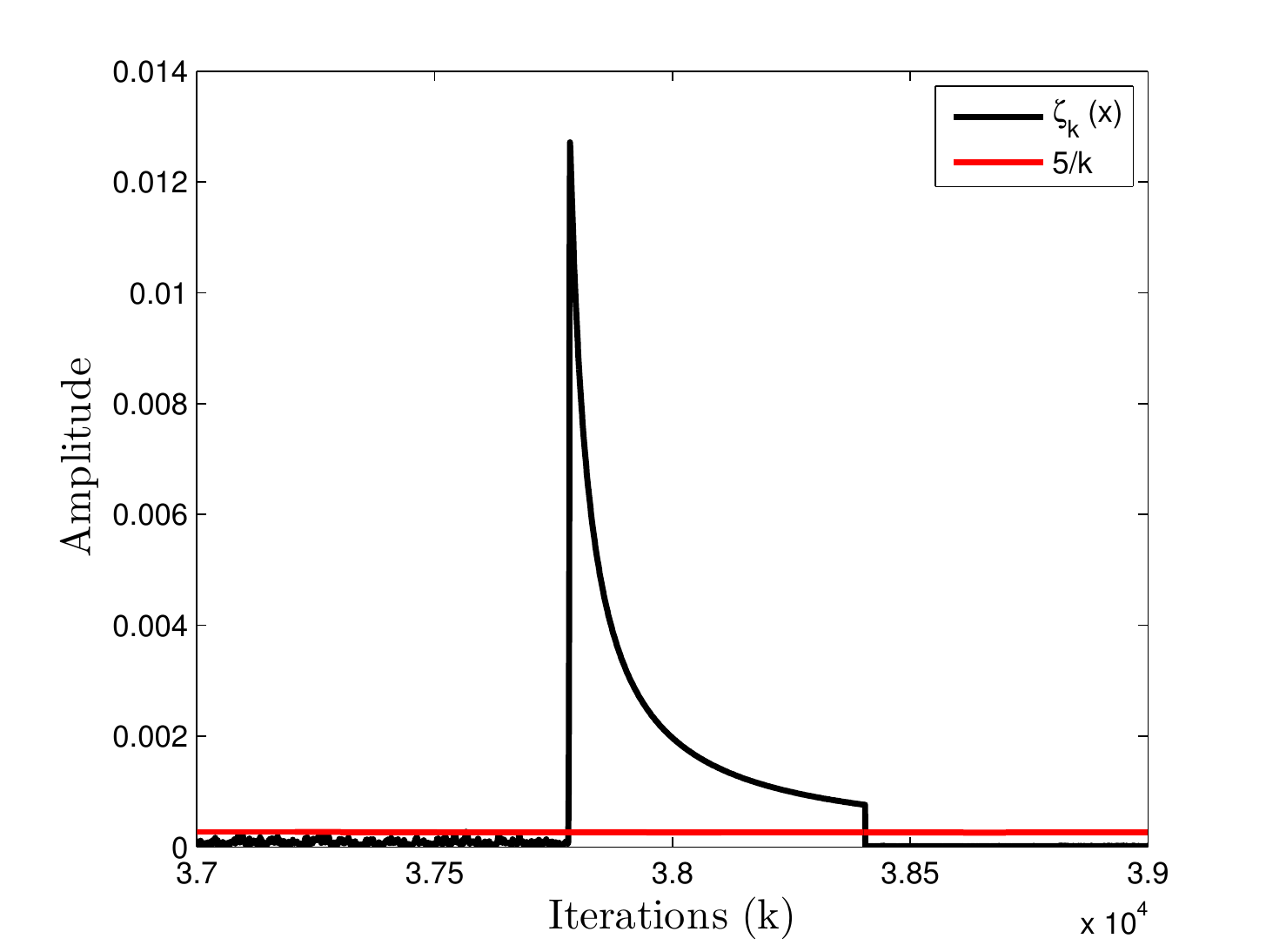}  
   \caption{Fault item 7 - normalized eccentricity  $\zeta_{k} (x)$ with $5/k $ $(m=3)$ threshold.}\label{fig:Falha7_e}
\end{subfigure}
\caption{Detection of outliers in the dataset: Behavior of fault item 7.}\label{fig:Falha7}
\end{figure}

In Figure \ref{fig:Falha7} it is possible to observe the results obtained for the item 7 signal, from Table \ref{tab:iten}. As within Figure \ref{fig:Falha1}, Figures \ref{fig:Falha7_v} illustrates the behavior of two elements of input $\bm{x}_{k}=[x^1_k \ x^2_k]$ in hardware architecture and in Figure \ref{fig:Falha7_e} it is possible to observe that there is a change of eccentricity (black curve), surpassing the value of the comparison threshold (red curve) also to $m=3$. The failure happens between moments $k=37700$ and $k=38400$.

Validation results in hardware architecture were compared with results obtained in a python software implementation of the algorithm TEDA. The hardware architecture was designed with floating point number format.

\subsection{Synthesis results} \label{sub:TEDAsintese}

After performing to validate the implemented circuit, the hardware synthesis was performed to obtain the FPGA resource occupation report, as well as the critical time information used to calculate the proposed implementation processing time. The floating point synthesis results were obtained for a Xilinx Virtex 6 xc6vlx240t-1ff1156 FPGA.

\subsubsection{Hardware occupation}

Table \ref{area_occupation} presents data related to the hardware occupation of the circuit implemented in the target FPGA. The first column shows the number of multipliers used, the second column displays the number of registers, and the third column shows the number of logical cells used as LUT ($n_{LUT}$) throughout the circuit.

\begin{table}[ht]
\centering
\caption{Hardware occupation.}
\label{area_occupation}
\begin{tabular}{cccc}
\hline
\textbf{Multipliers} & \begin{tabular}[c]{@{}c@{}}\textbf{Registers}\\ \end{tabular} & \textbf{$n_{LUT}$} \\ 
\hline
$27$ ($3$\%) & $414$ ($<1$\%) & $11.567$ ($7$\%) \\ 
\hline
\end{tabular}
\end{table}

Analyzing the data presented in Table \ref{area_occupation} it can be seen that even using a floating point resolution, which demands a greater amount of hardware resources than a fixed point implementation, only a small portion of the resources were occupied from the target FPGA, with a total of only about $3\%$ from multipliers, less than $1\%$ from registers, and about $7\%$ from logical cells used as LUT. With this, we found that the proposed circuit could also be applied in low cost FPGAs, where the amount of available hardware resources is even smaller. In addition, multiple TEDA modules could be applied in parallel for anomaly detection in the same dataset, in order to further reduce processing time.

\subsubsection{Processing time}

Table \ref{processsamento} presents information about the processing time (from Line \ref{alg:primeiraamostra} to Line \ref{alg_outlier4} in Algorithm \ref{alg_teda}) of the architecture implemented for the TEDA technique. The first column indicates the circuit critical time, $t_c$, the second column shows the initial delay, expressed by Equation \ref{delay}, the third column the TEDA run-time, expressed by Equation \ref{tTEDA}, and the last column the implementation throughput in samples per second (SPS), expressed by Equation \ref{thTEDA}, which consists of the amount of samples processed and classified (as normal or outlier) by TEDA every second.

\begin{table}[ht]
\centering
\caption{Processing time.}
\label{processsamento}
\begin{tabular}{cccc}
\hline
\textbf{Critical time} & \begin{tabular}[c]{@{}c@{}}\textbf{Delay}\\ \end{tabular} & \textbf{TEDA time} & \textbf{Throughput}\ \\ \hline
$138 \, \text{ns}$ & $414 \, \text{ns}$ & $138 \, \text{ns}$ & $7.2$ MSPS\\ \hline
\end{tabular}
\end{table}

The data presented in Table \ref{processsamento} are quite expressive. The circuit critical time, which also corresponds in the TEDA run-time, was only $t_c = 138 \, \text{ns}$. Thus, after the $414 \, \text{ns}$ delay, it is possible to get output for a processed sample sorted every $138 \, \text{ns}$, which guarantees a throughput of $7.2$ million sorted samples per second. These results indicate the feasibility of using the proposal presented in this work to manipulate large data flows in real time.

\subsection{Platforms comparison} \label{TEDAplatforms}

To date, no previous literature has been found to explore TEDA hardware implementations. Thus, this paper presents, for the first time, a proposal to implement the TEDA technique on FPGA. To verify the advantages of the hardware application proposed here over implementations on other software platforms, some comparisons of the FPGA processing time with the processing time of other software implementations were made. Table \ref{tab:comparacao} presents the results of the comparisons made. The first column indicates the hardware used, the second presents the processing time required to obtain the classification of each sample, and the third column, the speedup achieved by the proposal presented in this paper.

\begin{table}[ht]
\begin{center}
    \caption{Software implementations comparison.}\label{tab:comparacao}
\begin{tabular}{lcc}
\hline
\textbf{Platform}  & \textbf{Time} & \textbf{Speedup} \\ 
\hline
This work proposal on FPGA & $138 \, \text{ns}$  & $-$ \\ 
Python (Colab without GPU) & $435 \, \text{ms}$  & $3\text{,}000\text{,}000 \times$       \\ 
Python (Colab with Tesla K80 GPU) & $39.2 \, \text{ms}$  & $280\text{,}000 \times$         \\ 
Python (Local execution with 940 MX GPU) & $23.1 \, \text{ms}$ & $167\text{,}000 \times$         \\ 
\hline
\end{tabular}
\end{center}
\end{table}

The data presented in Table \ref{tab:comparacao} reaffirm the importance of this work. The hardware implementation on FPGA proposed here has been able to achieve speedups of up to $3$ million times compared to a Pyhton TEDA implementation using the Colab tool (without GPU processing). For the same Pyhton implementation using the Tesla K80 GPU processing Colab tool, a speedup of $280$ thousand times was obtained. In addition, when compared to a Python implementation on Intel(R) Core(TM) i7-7500U CPU@2.70GHz with 16 GB of RAM and GeForce 940 MX GPU, the hardware implementation on FPGA still had a $167$ thousand times advantage. Results that prove the advantages of using the proposal presented in this work to accelerate the TEDA technique, through the implementation on FPGA.

\section{Conclusion} \label{Sec:TEDAconclusion}
This work presented a proposal for hardware implementation of the TEDA data streaming anomaly detection technique. The hardware was implemented in RTL using floating point format. Synthesis results were obtained for a Xilinx Virtex 6 xc6vlx240t-1ff1156 FPGA. The proposed implementation used a small portion of the target FPGA resources, besides allowing the results to be obtained in a short processing time. The high speedups obtained in comparison with other software platforms reaffirmed the importance of this work, which is pioneering the hardware implementation of the TEDA technique on FPGA. The proposed architecture is feasible to be used in practical fault detection applications in real industrial processes with severe time constraints, as well as to handle large data volumes, such as data streaming, using low processing time.

\bibliography{PaperMain}

\end{document}